\documentclass[english,aps,manuscript,preprintnumbers]{revtex4-1}
\usepackage[T1]{fontenc}
\usepackage[latin9]{inputenc}
\usepackage{amsmath}
\usepackage{amssymb}
\usepackage{graphicx}
\usepackage{esint}
\usepackage{babel}
\begin{document}

\preprint{BROWN-HET-1706}

\title{dS/CFT and the operator product expansion}

\author{Atreya Chatterjee}
\email{atreya_chatterjee@brown.edu}

\author{David A. Lowe}
\email{lowe@brown.edu}

\affiliation{Department of Physics, Brown University, Providence, RI 02912, USA}
\begin{abstract}
Global conformal invariance determines the form of two and three-point
functions of quasi-primary operators in a conformal field theory,
and generates nontrivial relations between terms in the operator product
expansion. These ideas are generalized to the principal and complementary
series representations, which play an important role in the conjectured
dS/CFT correspondence. The conformal partial wave expansions are constructed
for these representations which in turn determine the operator product
expansion. This leads us to conclude that conformal field theories
containing such representations have essential singularities, so cannot
be realized as conventional field theories.
\end{abstract}
\maketitle

\section{Introduction}

The success of the anti-de Sitter/conformal field theory correspondence
(AdS/CFT) has inspired applications to de Sitter spacetime (dS) \citep{Strominger:2001pn}.
This leads one to try to find conformal field theories of relevance
to this correspondence which appear to exhibit novel properties, and
many have questioned whether such theories can be defined at all.

In the context of AdS/CFT detailed dictionaries relating the bulk
and boundary variables \citep{Aharony:1999ti} were found at the free
level. These ideas were generalized to a precision boundary/bulk correspondence,
order by order in a $1/N$ expansion in HKLL \citep{Hamilton:2006fh,Hamilton:2007wj,Lowe:2007ek,Kabat:2011rz}.
Our present goal is to attempt to extend such ideas to the dS/CFT
correspondence.

However, it has been difficult to produce examples in Minkowski space
and De-Sitter space because of various issues. Instability of string
theory in de-Sitter background \citep{Kachru:2003aw}, and the compact
spacelike boundary has made holography challenging in de-Sitter \citep{Witten:2001kn}.
We have discussed many of these problems, in our previous papers \citep{Chatterjee:2015pha,Chatterjee:2015wva}.
There we have shown how to extend the HKLL dictionary to dS space
for the case of non-interacting bulk theory. Many people have contributed
in understanding dS-holography including higher-spin holography for
dS\citep{Strominger:2001pn,Anninos:2011ui,Anninos:2013rza,Balasubramanian:2002zh,Banerjee:2013mca,Harlow:2011ke,Sarkar:2014dma,Spradlin:2001pw,Xiao:2014uea,Kastor:2002fu,Castro:2012gc}.

In section \ref{sec:Principal and Discrete series representations of bulk states.}
we introduce principal series and discrete series representation and
give a short description of the earlier work in maths literature.
Then we show how the generators act on the bulk fields. We derive
the bulk fields by solving the appropriate wave equation. Section
\ref{sec:Massless-scalar-field}, is devoted to the massless scalar
field. We find that modes of the massless scalar include both the
discrete series and a limit of the complementary series, which is
an indecomposable representation of the conformal group. This work
makes contact with recent work by Ashtekar et al. \citep{Ashtekar:2014zfa,Ashtekar:2015lla}
on the asymptotic boundary conditions in de Sitter spacetime. In particular
the discrete series modes carry vanishing energy, while the indecomposable
mode can carry energy, but changes the conformal structure of the
boundary. Both sets of operators are needed in the CFT to reproduce
a complete set of bulk modes.

It is the main goal of the present paper to construct the operator
product expansion in the conformal field theory for operators dual
to massive modes in the bulk. As is usual in conformal field theory,
the two and three-point functions of quasi-primary operators are determined
by conformal invariance. However when we explore the implications
of this for the operator product expansion, some surprising results
emerge, including the fact that the expansion involves terms with
arbitrarily rapid short distance singularities determined by a seemingly
infinite number of free parameters. This is in constrast to the more
ordinary CFTs appearing in the AdS/CFT correspondence, where the most
singular terms in the operator product expansion are determined by
the weights of the operators, and conformal invariance implies a single
parameter determines the full set of descendent couplings via conformal
partial waves. This leads us to conclude that such conformal field
theories do not exist in the space of ordinary renormalizable quantum
field theories, but rather share many of the features of non-renormalizable
field theories. For concreteness, many of our results are stated for
three-dimensional de Sitter spacetime. However since we only use the
global conformal group, the results are easily generalized to higher
dimensions.

\section{Principal and Discrete series representations of bulk states.\label{sec:Principal and Discrete series representations of bulk states.} }

The isometries of 3-dimensional dS form the group $SO(1,3)$. This
spacetime may be viewed as a hyperboloid embedded in 4-dimensional
Minkowski spacetime. The generators are given by $J_{i},K_{i}$ for
$i=1,2,3$. $J_{i}$ are the generators of rotation mixing three spacelike
embedding dimensions. $K_{i}$ are the generators of boost mixing
three spacelike dimension with the timelike dimension. There are various
Cartan sub-algebras of $SO(1,3)$. Depending on which Cartan subgroup
we choose, we get a different basis for the representations. One can
choose $SO(3)=\{J_{i}\}$ as the Cartan subgroup. Most papers in 1950-70
by Naimark, Tagirov, Chernikov, Raczka et al\citep{Chernikov:1968zm,Tagirov:1972vv,Raczka:discrete,Raczka:continuous,naimark64}
do that. So mode functions were labelled by quantum numbers $l,m$
(Eigenvalue of $\{J^{2},J_{3}\}$ respectively). $SO(3)$ (compact
group) has only finite dimensional representations $l=0,\frac{1}{2},1,\frac{3}{2},2,...,m=-l,-l+1,...,l$.
So range of $m$ is bounded for a given $l$.

On the CFT side, states are usually chosen as eigenstates of the $SU(1,1)$
Cartan sub-group. So it is useful to write the bulk generators $SO(1,3)$
as $SU_{L}(1,1)\otimes SU_{R}(1,1)$. (Just like $SL(2,C)\cong SU(1,1)\otimes SU(1,1)$.)
Combine the generators in the following way 
\begin{eqnarray*}
K_{1L}=\frac{1}{2}\left(-K_{1}+iJ_{1}\right)\ \ \ \ \  & K_{2L}=\frac{1}{2}\left(-K_{2}+iJ_{2}\right)\ \ \ \ \  & J_{L}=\frac{1}{2}\left(J_{3}+iK_{3}\right)\\
K_{1R}=\frac{1}{2}\left(K_{1}+iJ_{1}\right)\ \ \ \ \  & K_{2R}=\frac{1}{2}\left(K_{2}+iJ_{2}\right)\ \ \ \ \  & J_{R}=\frac{1}{2}\left(J_{3}-iK_{3}\right)\,.
\end{eqnarray*}
 Then 
\begin{eqnarray*}
[J_{L(R)},K_{1L(R)}] & = & iK_{2L(R)}\\
{}[J_{L(R)},K_{2L(R)}] & = & -iK_{1L(R)}\\
{}[K_{1L(R)},K_{2L(R)}] & = & -iJ_{L(R)}\,.
\end{eqnarray*}
Left and right sectors commute. We can also form the raising and lowering
operators $K_{\pm L(R)}=K_{1L(R)}\pm iK_{2L(R)}$. 
\begin{eqnarray*}
[J_{R},K_{\pm R}]=\pm K_{\pm R} & \ \ \ \ \  & [J_{L},K_{\pm L}]=\pm K_{\pm L}\,.
\end{eqnarray*}
Thus $\{J_{L(R)},K_{\pm L(R)}\}$ form $SU_{L(R)}(1,1)$ group.

Now let us discuss unitary irreducible representations of $SU(1,1)$.
States are labelled by eigenvalues of $\{C_{L}=J_{L}^{2}-K_{1L}^{2}-K_{2L}^{2},J_{L},J_{R}\}$
\begin{eqnarray*}
C_{L}|h,l\rangle & = & h(h-1)|h,l\rangle\\
J_{L}|h,l\rangle & = & l|h,l\rangle\,.
\end{eqnarray*}

Irreducible representations split into discrete series and continuous
series (principal and complementary series) \citep{Raczka:discrete,Raczka:continuous,Lindbad:cont:su11,Husz:horospheric_basis,Mukunda:O11,Mukunda:O21,Lindbad:eigenfunction,Parentini:discrete,Parentini:principal}.
In the discrete series $h=-n/2,n\in N$. $l=-h,-h+1,...$ for positive
discrete series $D^{+}$(lowest weight) and $l=h,h-1,...$ for negative
discrete series $D^{-}$(highest weight). For continuous series $h=-\frac{1}{2}+i\rho,\,0<\rho<\infty$
and $l=0,\pm1,\pm2,...$ or $l=\pm\frac{1}{2},\pm\frac{3}{2},...$
corresponding to $C_{\rho}^{0}$ or $C_{\rho}^{1/2}$ respectively.

Similarly, $SU_{R}(1,1)$ sector can be constructed. For scalar fields
$h_{L}=h_{R}=h$. Casimir of $SO(1,3)$ is then given by
\begin{eqnarray*}
C & = & C_{L}+C_{R}\\
 & = & 2h(h-1)\,.
\end{eqnarray*}
For the discrete series 
\begin{eqnarray*}
C & = & -\frac{1}{2}n(2-n),\,n\in N\,.
\end{eqnarray*}
For the continuous series
\begin{eqnarray*}
C & = & -2\rho^{2}-\frac{1}{2},\,0<\rho<\infty\,.
\end{eqnarray*}
As we will see, some modes of the massless scalar correspond to the
$n=2$ discrete series. There $l=\pm1,\pm2,...$ for $D^{\pm}$ respectively. 

\subsection{Action of the generators on the states}

\begin{figure}
\centering \includegraphics[scale=0.7]{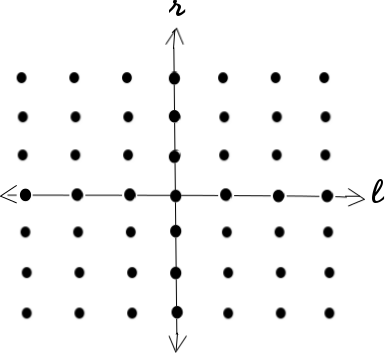}\label{principal series}
\caption{Weight space diagram for principal series. x-y axes are the $l,r$
values. Solid dots represent states for all $l,r\in\mathbb{{Z}}$.
These states have both growing and decaying modes. $K_{\pm L}$ shift
the states right and left respectively. Similarly $K_{\pm R}$ shift
the states up and down respectively.}

\label{fig:principal series} 
\end{figure}

\begin{figure}
\centering \includegraphics[scale=0.7]{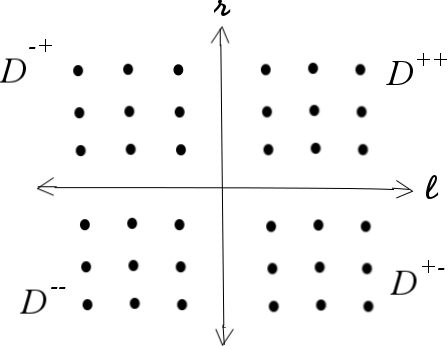}\label{discrete series}
\caption{Weight space diagram for discrete series x-y axes are the $l,r$ values.
Solid dots represent states for all non zero $l,r$. These states
contain only decaying modes. $K_{\pm L}$ shift the states right and
left respectively. $K_{\pm L}$ annihilates $l=\mp1$ states respectively.
Similarly $K_{\pm R}$ shift the states up and down respectively.
$K_{\pm R}$ annihilates $r=\mp1$ states respectively.}

\label{fig:discrete series} 
\end{figure}

Let us write below action of all the generators on the state $|h,l,r\rangle$
\begin{eqnarray}
J_{R}|h,l,r\rangle & = & r|h,l,r\rangle\label{eq:Jr}\\
J_{L}|h,l,r\rangle & = & l|h,l,r\rangle\label{eq:Jl}\\
C_{L}|h,l,r\rangle & = & h\left(h-1\right)|h,l,r\rangle\nonumber \\
C_{R}|h,l,r\rangle & = & h\left(h-1\right)|h,l,r\rangle\nonumber \\
C|h,l,r\rangle & = & \left(C_{L}+C_{R}\right)|h,l,r\rangle=2h\left(h-1\right)|h,l,r\rangle\label{eq:C}\\
K_{\pm L}|h,l,r\rangle & = & i\left(\pm\left(h-1\right)-l\right)|j_{L},l\pm1,r\rangle\\
K_{\pm R}|h,l,r\rangle & = & i\left(\mp\left(h-1\right)-r\right)|j_{L},l,r\pm1\rangle\,.
\end{eqnarray}
For a scalar field of mass $m$, $4h\left(h-1\right)=m^{2}\implies h_{\pm}=\frac{1}{2}\pm\sqrt{\frac{1}{4}-\frac{m^{2}}{4}}$
and $l(r)=0,\pm1,\pm2,...$ or $l(r)=\pm\frac{1}{2},\pm\frac{3}{2},...$.
The principal series corresponds to $m>1$ and the complementary series
corresponds to $1>m>0$. A component of the massless scalar behaves
like a discrete series with $h=1$. Figure \ref{fig:principal series}
and \ref{fig:discrete series} show the weight space diagram for principal
series and discrete series. Similar weight space diagrams for representation
in Anti-de Sitter space was given by Dusedau and Freedman\citep{Dusedau:1985ue}.

\subsection{States in coordinate space}

Now we know how the generators act on the states. To explore bulk-boundary
correspondence, we want to see how the states behave close to the
boundary. It is convenient to transform to a basis of eigenstates
in coordinate space.

De Sitter space can be described by the flat slicing coordinates $\eta,z,\bar{z}$
\begin{eqnarray*}
ds^{2} & = & \frac{1}{\eta^{2}}\left(-d\eta^{2}+dzd\bar{z}\right)\,.
\end{eqnarray*}
There are many nice reviews of de-Sitter space \citep{Spradlin:2001pw}.
$z$ is complexified spacelike coordinate. $\eta$ is timelike coordinate.
De-Sitter has boundary at future and past infinity $\eta\rightarrow0$.
Bulk isometry generators are
\begin{eqnarray*}
J_{L}=z\partial_{z}+\frac{\eta}{2}\partial_{\eta}\ \ \ ,\ \ \  & K_{+L}=i\left(z^{2}\partial_{z}+\eta^{2}\partial_{\bar{z}}+z\eta\partial_{\eta}\right)\ \ \ ,\ \ \  & K_{-L}=-i\partial_{z}\\
J_{R}=-\bar{z}\partial_{\bar{z}}-\frac{\eta}{2}\partial_{\eta}\ \ \ ,\ \ \  & K_{-R}=-i\left(\bar{z}^{2}\partial_{\bar{z}}+\eta^{2}\partial_{z}+\bar{z}\eta\partial_{\eta}\right)\ \ \ ,\ \ \  & K_{+R}=i\partial_{\bar{z}}\,.
\end{eqnarray*}
Note that if we put $\eta\rightarrow0$ and $\eta\partial_{\eta}\rightarrow2h$
as we approach the boundary then
\begin{eqnarray*}
J_{L}\rightarrow-L_{0}\ \ \ \ \  & K_{+L}\rightarrow-iL_{1}\ \ \ \ \  & K_{-L}\rightarrow iL_{-1}\\
J_{R}\rightarrow\bar{L}_{0}\ \ \ \ \  & K_{+R}\rightarrow-i\bar{L}_{-1}\ \ \ \ \  & K_{-R}\rightarrow i\bar{L}_{1}
\end{eqnarray*}
as shown in the appendix. Casimir operator is given by
\begin{eqnarray*}
C & = & C_{L}+C_{R}\,.
\end{eqnarray*}

Simultaneous eigenstates of $J_{L},J_{R},C$ with eigenvalues $l,r,\frac{m^{2}}{4}$
($m$ is mass) respectively, form the principal series representation.
Solving the differential equations \eqref{eq:Jr}, \eqref{eq:Jl}
and \eqref{eq:C} we get 
\begin{eqnarray}
\phi_{l,r}(z,\bar{z},\eta) & = & \left(\frac{z}{\bar{z}}\right)^{\frac{l+r}{2}}\eta^{l-r}\Big[A_{1}i^{-l-r}\left(\frac{z\bar{z}}{\eta^{2}}\right)^{-\frac{l+r}{2}}\ _{2}F_{1}\left(\frac{1-2l-\sqrt{1-m^{2}}}{2},\frac{1-2l+\sqrt{1-m^{2}}}{2},1-l-r,\frac{z\bar{z}}{\eta^{2}}\right)\nonumber \\
 &  & +A_{2}i^{l+r}\left(\frac{z\bar{z}}{\eta^{2}}\right)^{\frac{l+r}{2}}\ _{2}F_{1}\left(\frac{1+2r-\sqrt{1-m^{2}}}{2},\frac{1+2r+\sqrt{1-m^{2}}}{2},1+l+r,\frac{z\bar{z}}{\eta^{2}}\right)\Big]\,.\label{eq:general solution}
\end{eqnarray}

Near the boundary ($\eta\rightarrow0$) it behaves like
\begin{eqnarray*}
\phi_{l,r}(z,\bar{z},\eta\rightarrow0) & = & \left(\frac{z}{\bar{z}}\right)^{\frac{l+r}{2}}\eta^{l-r}\Big[A_{1}i^{-l-r}\left(\frac{z\bar{z}}{\eta^{2}}\right)^{-\frac{l+r}{2}}\left(a_{1}\left(\frac{\eta^{2}}{z\bar{z}}\right)^{-l+h_{-}}+a_{2}\left(\frac{\eta^{2}}{z\bar{z}}\right)^{-l+h_{+}}\right)\\
 &  & +A_{2}i^{l+r}\left(\frac{z\bar{z}}{\eta^{2}}\right)^{\frac{l+r}{2}}\left(a_{1}\left(\frac{\eta^{2}}{z\bar{z}}\right)^{r+h_{-}}+a_{2}\left(\frac{\eta^{2}}{z\bar{z}}\right)^{r+h_{+}}\right)\Big]\\
 & = & b_{-}\eta^{2h_{-}}\left(\frac{1}{z^{h_{-}-l}\bar{z}^{h_{-}+r}}\right)+b_{+}\eta^{2h_{+}}\left(\frac{1}{z^{h_{+}-l}\bar{z}^{h_{+}+r}}\right)\\
 & = & b_{-}\eta^{2h_{-}}O_{l,r,h_{-}}(z,\bar{z})+b_{+}\eta^{2h_{+}}O_{l,r,h_{+}}(z,\bar{z})
\end{eqnarray*}
where $b_{\pm}$ are some constants and $h_{\pm}=\frac{1\pm\sqrt{1-m^{2}}}{2}$.
Here $-\infty\leq l,r\leq\infty.$ Another important thing to note
is that $\phi_{l,r}\sim z^{l}\bar{z}^{r}$ (power law).
\begin{eqnarray*}
\phi_{l,r}(z,\bar{z},\eta\rightarrow0) & = & \left(b_{-}\left(\frac{\eta^{2}}{z\bar{z}}\right)^{h_{-}}+b_{+}\left(\frac{\eta^{2}}{z\bar{z}}\right)^{h_{+}}\right)z^{l}\bar{z}^{-r}\,.
\end{eqnarray*}

For principal series $h_{-},h_{+}$ are complex conjugate of each
other. So the modes oscillate close to the boundary. For complementary
series $h_{-}<0<h_{+}$ and real. So half of the modes grow $\left(\eta^{2h_{-}}\right)$
and other half of the modes decay $\left(\eta^{2h_{+}}\right)$ near
the boundary. They are respectively called growing and decaying mode. 

\section{Massless scalar field\label{sec:Massless-scalar-field}}

Now we are going to look into the massless case. There are two ways
that representations contribute to the massless scalar. 

\subsection{Limit of complementary series}

One is the $m\rightarrow0$ limit of equation \eqref{eq:general solution}.
This is the limit of complementary series representation. 
\begin{eqnarray}
\phi_{l,r}(z,\bar{z},\eta) & = & \left(\frac{z}{\bar{z}}\right)^{\frac{l+r}{2}}\eta^{l-r}\Big[A_{1}i^{-l-r}\left(\frac{z\bar{z}}{\eta^{2}}\right)^{-\frac{l+r}{2}}\ _{2}F_{1}\left(-l,1-l,1-l-r,\frac{z\bar{z}}{\eta^{2}}\right)\nonumber \\
 &  & +A_{2}i^{l+r}\left(\frac{z\bar{z}}{\eta^{2}}\right)^{\frac{l+r}{2}}\ _{2}F_{1}\left(r,1+r,1+l+r,\frac{z\bar{z}}{\eta^{2}}\right)\Big]\,.\label{eq:limit comp}
\end{eqnarray}
Close to the boundary it goes like
\begin{eqnarray*}
\phi_{l,r}(z,\bar{z},\eta\rightarrow0) & = & \left(b_{-}+b_{+}\frac{\eta^{2}}{z\bar{z}}\right)z^{l}\bar{z}^{-r}\,.
\end{eqnarray*}
Note that, it has both the decaying mode and the constant mode. 

\subsection{Discrete series}

Second is the Discrete series representation. There are two ways of
deriving discrete series. Let us first see how it is derived in earlier
math papers \citep{Raczka:discrete,Tagirov:1972vv,Chernikov:1968zm}.
First find the eigenstates of $\Box|l,m\rangle=-m^{2}|l,m\rangle$
in the $|l,m\rangle$ basis (eigenstate of $\{J^{2},J_{3}\}$). $\Box$
is second order differential equation and we get two independent solutions.
Then choose only the decaying modes. This removes half of the solutions.
This condition results in discrete eigenvalues ($-m^{2}$) of $\Box$.
Hence the representation is called Discrete series. Note that this
is in agreement with the previous section where we said that for massless
discrete series $h=1$.

Now let us derive the discrete series in another way. Diagonalize
the Hilbert space in the eigenstates of $J_{L},J_{R}$. In addition
to equations \eqref{eq:Jr}, \eqref{eq:Jl} and \eqref{eq:C}(with
$h=1$) states have to satisfy equations
\begin{eqnarray}
C_{L}|1,l,r\rangle & = & 0\nonumber \\
C_{R}|1,l,r\rangle & = & 0\nonumber \\
K_{\pm L}|1,\mp1,r\rangle & = & 0\label{eq:highest l}\\
K_{\pm R}|1,l,\mp1\rangle & = & 0\,.\label{eq:highest r}
\end{eqnarray}

There are four sectors as shown in figure \ref{fig:discrete series}.
$D^{m_{L}m_{R}}$ where $l(r)=-1$ is the lowest weight and $l(r)=1$
is the highest weight state. Thus\eqref{eq:highest l} and \eqref{eq:highest r}.
In this basis, highest and lowest weight states are manifest. This
is over-constrained set of equations. Equation \eqref{eq:highest l}
and \eqref{eq:highest r} are first order differential equation which
has only one solution. As a result, half of the general solution of
equation \eqref{eq:C} is removed. We find that eigenstates decay
near boundary. To see this consider the following states
\begin{eqnarray*}
\phi_{-1,r}^{D}(z,\bar{z},\eta) & = & A\left(\frac{z}{\eta}\right)^{r-1}\left(\frac{z\bar{z}}{\eta}-\eta\right)^{-1-r}\\
\phi_{1,r}^{D}(z,\bar{z},\eta) & = & A\bar{z}{}^{-\frac{r+1}{2}}\eta^{2}\\
\phi_{l,1}^{D}(z,\bar{z},\eta) & = & A\left(\frac{\bar{z}}{\eta}\right)^{-l-1}\left(\frac{z\bar{z}}{\eta}-\eta\right)^{-1+l}\\
\phi_{l,-1}^{D}(z,\bar{z},\eta) & = & Az^{\frac{l-1}{2}}\eta^{2}\,.
\end{eqnarray*}

Note that close to boundary all the above solutions go like $\eta^{2}$.
All other states can be obtained by acting with $K_{\pm L},K_{\pm R}$.
Since $K_{\pm L},K_{\pm R}$ do not decrease the power of $\eta$,
all the states will have same $\eta$ dependence. Hence all the modes
of the discrete series decay near the boundary. This also shows that
$h=1$. This suggests that these states are a linear combination of
states found in the previous approach. 

So either imposing regularity of the modes in the $|m,l\rangle$ basis
is equivalent to requiring the existence of a highest weight state
in $|1,l,r\rangle$ basis. It removes the half of the modes which
stay constant near the boundary. On the other hand, the limit of the
complementary series has both growing and decaying modes.

\subsection{Discrete series cannot carry energy in dS. }

Now that we have understood discrete series and limit of complementary
series in more detail, what are the physical consequences? Does graviton
belong to discrete series or complementary series? Ashtekar et al.,
in a series of papers \citep{Ashtekar:2014zfa,Ashtekar:2015lla},
has shown that gravity waves in de-Sitter cannot carry energy if the
constant modes of the gravitons are removed. In light of this,
\begin{enumerate}
\item If the gravitons are described by discrete series then the constant
modes are absent. Then gravity waves cannot carry energy.
\item If we want graviton modes to carry energy and a complete set of modes,
the gravitons must contain modes from the limit of complementary series.
\end{enumerate}

\subsection{Indecomposiblity of limit of complementary series}

In this section we will show that limit of complementary series is
indecomposible. A representation is indecomposible\citep{Wybourne}
if it cannot be separated into two or more irreducible representations.
We have already shown that decaying modes form the irreducible discrete
series representation. Then the question is: Does the remaining constant
mode also form irreducible representation? 

To establish this we show that constant modes turn into decaying modes
under the action of generators. Schematically, 
\begin{eqnarray*}
K|decay\rangle & \rightarrow & |decay\rangle\\
K|constant\rangle_{m_{l}\neq0} & \rightarrow & |constant\rangle\\
K|constant\rangle_{m_{l}=0} & \rightarrow & |decay\rangle
\end{eqnarray*}
where $K$ is some ladder operator. Equation \eqref{eq:limit comp}
is the general solution of the massless scalar field. Schematically
the two independent solutions are 
\begin{eqnarray*}
\phi_{l,r}(z,\bar{z},\eta\rightarrow0) & = & \left(b_{-}+b_{+}\frac{\eta^{2}}{z\bar{z}}\right)z^{l}\bar{z}^{-r}\\
 & = & \left(b_{-}|constant\rangle+b_{+}\frac{|decay\rangle}{z\bar{z}}\right)z^{l}\bar{z}^{-r}\\
|decay\rangle_{\eta\rightarrow0} & = & \eta^{2}\\
|constant\rangle_{\eta\rightarrow0} & = & 1
\end{eqnarray*}
where $b_{-},b_{+}$ are some constants. $|decay\rangle$ modes form
the irreducible discrete series. They are either highest or lowest
weight representations. This we have discussed in previous section.

\begin{figure}
\centering \includegraphics[scale=0.7]{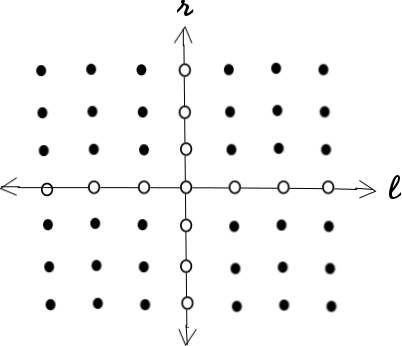}\label{limit complementary series}
\caption{Weight space diagram for limit of complementary series. x-y axes are
the $l,r$ values. Solid dots represent states for all non-zero $l,r$.
These states have both constant and decaying modes. Empty dots for
$l=0$ or $r=0$ represent states which have only constant modes.
$K_{\pm L}$ shift the states right and left respectively.$K_{\pm L}$
acting on constant modes of $l=0$ states, convert them to decaying
modes. Similarly $K_{\pm R}$ shift the states up and down respectively.
$K_{\pm R}$ acting on constant modes of $r=0$ states, convert them
to decaying modes.}

\label{fig:limit complementary series} 
\end{figure}

To understand the issue let us see the general $|-1,l=0,r\rangle$
mode
\begin{eqnarray*}
\phi_{0,r}(z,\bar{z},\eta) & = & \left(\frac{z}{\bar{z}}\right)^{\frac{r}{2}}\eta^{-r}\Big[A_{1}i^{-r}\left(\frac{z\bar{z}}{\eta^{2}}\right)^{-\frac{r}{2}}\ _{2}F_{1}\left(0,1,1-r,\frac{z\bar{z}}{\eta^{2}}\right)\\
 &  & +A_{2}i^{r}\left(\frac{z\bar{z}}{\eta^{2}}\right)^{\frac{r}{2}}\ _{2}F_{1}\left(r,1+r,1+r,\frac{z\bar{z}}{\eta^{2}}\right)\Big]\\
 & = & A_{1}i^{-r}\bar{z}^{-r}+A_{2}i^{r}\left(\frac{z}{\eta^{2}}\right)^{r}\left(1-\frac{z\bar{z}}{\eta^{2}}\right)^{-r}\,.
\end{eqnarray*}
We see that there is only a constant part. Now let us apply $K_{-L},K_{+L}$
\begin{eqnarray*}
K_{-L}|0,r\rangle & = & A_{2}ri^{-1+r}z^{-1+r}\eta^{2r}\left(1-\frac{z\bar{z}}{\eta^{2}}\right)^{-r-1}=|-1,r\rangle_{decay}\\
K_{+L}|0,r\rangle & = & A_{1}ri^{-1-r}\bar{z}^{-1-r}\eta^{2}=|1,r\rangle_{decay}\,.
\end{eqnarray*}
Thus we get only the decaying modes. This shows that the growing modes
convert into decaying modes and proves that limit of complementary
series is indecomposible representation. Figure \ref{fig:limit complementary series}
gives the weight space diagram for the limit of complementary series
to illustrate this point.

\section{Transformation from $|l,r\rangle$ basis to momentum basis\label{sec:Transformation-from-}}

In this section we derive the transformation from a momentum basis
(eigenstate of $L_{-1}$ operator) to the $l,r$ basis (eigenstates
of $L_{0},\bar{L}_{0}$ operator with eigenvalue $l,r$ respectively.
). One reason to do it is that the scalar field in the bulk is generally
written in momentum basis but boundary operators are generally expressed
in $l,r$ basis. Subsection \eqref{subsec:Principal-series} gives
in detail the calculations for principal series. In subsection \eqref{subsec:Summary-(comparing-principal}
we summarize the main results and compare the differences between
the two representations. 

\subsection{Principal series\label{subsec:Principal-series}}

Momentum basis are eigenstates of $L_{-1}=-\partial_{z},\bar{L}_{-1}=-\partial_{\bar{z}}$.
$l,r$ basis are eigenstate of $L_{0}=-\left(z\partial_{z}+h\right),\bar{L}_{0}=-\left(\bar{z}\partial_{\bar{z}}+\bar{h}\right)$
respectively. We want to find the coefficients $c_{k,l,r}$ of the
relation 
\begin{eqnarray}
|k,\bar{k}\rangle & = & \sum c_{k,l,r}|l,r\rangle\label{eq:plr}
\end{eqnarray}

Our approach is similar to what Lindbad et al do in section (4A) of
\citep{Lindbad:cont:su11}. From commutation relation $[L_{n},\phi_{l}]=\left(\left(h-1\right)n-l\right)\phi_{n+l}$
we get 
\begin{eqnarray}
L_{0}|l,r\rangle & = & -l|l,r\rangle\label{eq:prin-gen action 1}\\
L_{1}|l,r\rangle & = & \left(h-1-l\right)|l+1,r\rangle\label{eq:prin-gen action 4}\\
L_{-1}|l,r\rangle & = & \left(1-h-l\right)|l-1,r\rangle\label{eq:prin-gen action 2}\\
L_{-1}|k,\bar{k}\rangle & = & \left(\frac{i\bar{k}}{2}\right)|k,\bar{k}\rangle\,.\label{eq:prin-gen action 3}
\end{eqnarray}
To find $c_{k,l,r}$ we act with $L_{-1}$ on both the side of equation
\eqref{eq:plr} 
\begin{eqnarray*}
L_{-1}|k,\bar{k}\rangle & = & \sum c_{k,l,r}L_{-1}|l,r\rangle\\
\left(\frac{i\bar{k}}{2}\right)\langle l',r'|k,\bar{k}\rangle & = & \sum c_{k,l,r}(-l+1-h)\langle l',r'|l-1,r\rangle\\
\left(\frac{i\bar{k}}{2}\right)c_{k,l',r'} & = & c_{k,l'+1,r'}(-l'-h)\,.
\end{eqnarray*}
Solving the recurrence relation we get 
\begin{eqnarray*}
c_{k,l,r} & = & c_{k,l-1,r}\left(\frac{-i\bar{k}}{2}\right)\frac{1}{(h+l-1)}\\
c_{k,l,r} & = & \left(\frac{-i\bar{k}}{2}\right)^{l}\frac{h!}{(h+l-1)!}c_{k,0,r}\\
 & = & \left(\frac{-i\bar{k}}{2}\right)^{l}\frac{\Gamma(h+1)}{\Gamma(h+l)}c_{k,0,r}\\
 & = & \left(\frac{-i\bar{k}}{2}\right)^{l}\frac{\sin\pi h}{\pi}(-1)^{l}\Gamma(h+1)\Gamma(1-h-l)c_{k,0,r}\,.
\end{eqnarray*}

Similarly one can derive $c_{k,0,r}$ by the action of $\bar{L}_{-1}$.
Finally one gets
\begin{eqnarray*}
c_{k,l,r} & = & \left(\frac{ik}{2}\right)^{r}\left(\frac{i\bar{k}}{2}\right)^{l}\left(\frac{\sin\pi h}{\pi}\right)^{2}\Gamma(h+1)^{2}\Gamma(1-h-l)\Gamma(1-h-r)c_{k,0,0}\,.
\end{eqnarray*}

We choose normalization $c_{k,0,0}=\frac{\pi^{2}}{(\sin\pi h)^{2}(-2i)^{2h+1}\Gamma(h+1)^{2}|ik/2|}$.
Plugging this back into equation \eqref{eq:plr} we get 
\begin{eqnarray}
|k,\bar{k}\rangle & = & \sum\left(\frac{ik}{2}\right)^{r-1/2}\left(\frac{i\bar{k}}{2}\right)^{l-1/2}(-2i)^{-2h-1}\Gamma(1-h-l)\Gamma(1-h-r)|l,r\rangle\,.\label{eq:k-l}
\end{eqnarray}

We can now invert equation \eqref{eq:k-l}.
\begin{eqnarray}
|l,r\rangle & = & \frac{1}{(2\pi i)^{2}}\oint dkd\bar{k}\left(\frac{ik}{2}\right)^{-r-1/2}\left(\frac{i\bar{k}}{2}\right)^{-l-1/2}\frac{(-2)^{2h+1}}{\Gamma(1-h-l)\Gamma(1-h-r)}|k,\bar{k}\rangle\,.\label{eq:l-k}
\end{eqnarray}

One can check that this is consistent with equation \eqref{eq:k-l}.
To see that start with the RHS of the above equation, substitute $|k,\bar{k}\rangle$
from equation \eqref{eq:k-l} and we get the LHS of above equation
\begin{eqnarray*}
 &  & \frac{1}{(2\pi i)^{2}}\oint dkd\bar{k}\left(\frac{ik}{2}\right)^{-r'-1/2}\left(\frac{i\bar{k}}{2}\right)^{-l'-1/2}\frac{(-2i)^{2h-1}}{\Gamma(1-h-l')\Gamma(1-h-r')}|k,\bar{k}\rangle\\
 & = & \sum_{l.r}\frac{1}{(2\pi i)^{2}}\oint dkd\bar{k}\left(\frac{ik}{2}\right)^{-r'+r-1}\left(\frac{i\bar{k}}{2}\right)^{-l'+l-1}\frac{-\Gamma(1-h-l)\Gamma(1-h-r)}{4\Gamma(1-h-l')\Gamma(1-h-r')}|l,r\rangle\\
 & = & \sum_{l.r}\delta_{ll'}\delta_{rr'}\frac{\Gamma(1-h-l)\Gamma(1-h-r)}{\Gamma(1-h-l')\Gamma(1-h-r')}|l,r\rangle\\
 & = & |l',r'\rangle\,.
\end{eqnarray*}

Now we know the basis transformations each way $|k,\bar{k}\rangle\leftrightarrow|l,r\rangle$,
we can write this as a boundary operator/state correspondence as follows
\begin{eqnarray*}
O(z,\bar{z})|0\rangle=|z,\bar{z}\rangle & = & \sum_{l,r}\frac{1}{z^{h+l}\bar{z}^{h+r}}|l,r\rangle\\
O(z,\bar{z})|0\rangle=|z,\bar{z}\rangle & = & \oint\frac{dkd\bar{k}}{(2\pi i)^{2}}e^{\frac{i}{2}\left(k\bar{z}+\bar{k}z\right)}|k|^{2h-1}|k,\bar{k}\rangle\,.
\end{eqnarray*}

So RHS of the above two equations must be equal. That is 
\begin{eqnarray}
\sum_{l,r}\frac{1}{z^{h+r}\bar{z}^{h+l}}|l,r\rangle & = & \oint\frac{dkd\bar{k}}{(2\pi i)^{2}}e^{\frac{i}{2}\left(k\bar{z}+\bar{k}z\right)}|k|^{2h-1}|k,\bar{k}\rangle\,.\label{eq:z-l-k}
\end{eqnarray}
To verify that, substitute $|l,r\rangle$ from equation \eqref{eq:l-k}
in LHS to get the RHS. 
\begin{eqnarray*}
|z,\bar{z}\rangle & = & \sum_{l,r}\frac{1}{z^{h+l}\bar{z}^{h+r}}|l,r\rangle\\
 & = & \frac{1}{(2\pi i)^{2}}\oint dkd\bar{k}\sum_{l,r}\left(\frac{ik\bar{z}}{2}\right)^{-r-1/2}\left(\frac{i\bar{k}z}{2}\right)^{-l-1/2}\frac{(-2i)^{2h-1}}{\Gamma(1-h-l)\Gamma(1-h-r)}|k,\bar{k}\rangle\\
 & = & \frac{1}{(2\pi i)^{2}}\oint dkd\bar{k}|k|^{2h-1}\left(\sum_{l,r}\left(\frac{ik\bar{z}}{2}\right)^{-h-r}\left(\frac{i\bar{k}z}{2}\right)^{-h-l}\frac{1}{\Gamma(1-h-l)\Gamma(1-h-r)}\right)|k,\bar{k}\rangle\\
 & = & \frac{1}{(2\pi i)^{2}}\oint dkd\bar{k}|k|^{2h-1}e^{\frac{i}{2}\left(k\bar{z}+\bar{k}z\right)}|k,\bar{k}\rangle\,.
\end{eqnarray*}
Here we have used the identity 
\begin{eqnarray*}
e^{z} & = & \sum_{n\in Z}\frac{z^{h+n}}{\Gamma(h+n+1)}\,.
\end{eqnarray*}
When $h$ is integer, $\frac{1}{\Gamma(h+n+1)}=0$ for $n<-h$. Thus
\begin{eqnarray*}
e^{z} & = & \sum_{n\geq-h}\frac{z^{h+n}}{\Gamma(h+n+1)}\\
 & = & \sum_{m\geq0}\frac{z^{m}}{\Gamma(m+1)}
\end{eqnarray*}
coincides with the usual definition of exponential function. When
$h$ is non integer, negative powers of $z$ appear in the sum. Each
such term diverges at the the origin but the sum is finite. 

\subsection{Summary \label{subsec:Summary-(comparing-principal}}

The boundary operator/state correspondence is
\begin{eqnarray*}
O(z,\bar{z})|0\rangle & = & \begin{cases}
\sum_{l,r\leq0}\frac{1}{z^{l}\bar{z}^{r}}|l-h,r-h\rangle & \text{{Highest\,Weight}}\\
\sum_{l,r}\frac{1}{z^{h+l}\bar{z}^{h+r}}|l,r\rangle & \text{{Principal\,Series}}
\end{cases}
\end{eqnarray*}
Transforming to the momentum basis we get 
\begin{eqnarray*}
O(z,\bar{z})|0\rangle & = & \frac{1}{(2\pi i)^{2}}\oint dkd\bar{k}|k|^{2h-1}e^{\frac{i}{2}\left(k\bar{z}+\bar{k}z\right)}|k,\bar{k}\rangle\,.
\end{eqnarray*}
We can also transform back. As we have already stated in subsection
\eqref{subsec:Principal-series}, the key identity is
\begin{eqnarray*}
e^{z} & = & \sum_{n\in Z}\frac{z^{h+n}}{\Gamma(h+n+1)}=\sum_{m\geq0}\frac{z^{m}}{\Gamma(m+1)}
\end{eqnarray*}
where $h$ is integer. 

In the momentum basis, the expansion
\begin{eqnarray*}
O(z,\bar{z})|0\rangle & = & \oint\frac{dkd\bar{k}}{(2\pi i)^{2}}e^{\frac{i}{2}\left(k\bar{z}+\bar{k}z\right)}|k|^{2h-1}|k,\bar{k}\rangle
\end{eqnarray*}
takes the same form for both the highest weight and principal series
representation. So any two or three-point correlation function in
the momentum or position basis is going to have the same scaling form
for principal series and highest weight representation since the form
is fixed by conformal symmetry. For example,
\begin{eqnarray*}
\langle O(z)O(w)\rangle_{\text{{principal-series}}} & = & \frac{1}{\left(z-w\right)^{2h}}\\
\langle O(z)O(w)\rangle_{\text{{highest-weight}}} & = & \frac{1}{\left(z-w\right)^{2h}}
\end{eqnarray*}
where $h$ is the weight of operator. It is real for highest weight
rep but complex for principal series. Now consider the following 2-point
function $\langle\oint w^{k+h}\phi(w)\oint z^{k}\phi(z)\rangle$ ($h$
is the weight of the operator). For highest weight representation
\begin{eqnarray*}
\langle\oint w^{k+h}\phi_{H}(w)\oint z^{k}\phi_{H}(z)\rangle & = & 0\left(\text{{for} }k\in\mathbb{{Z}}_{+}\right)
\end{eqnarray*}
because $\oint z^{k}\phi_{H}(z)|0\rangle=0$ for $k\in\mathbb{{Z}}_{+}$. 

For principal series representation it gives
\begin{eqnarray*}
 &  & \langle\oint w^{k+h}\phi_{NH}(w)\oint z^{k}\phi_{NH}(z)\rangle\\
 & = & \langle\sum_{-\infty<m<\infty}\oint w^{k-m}dw\phi_{m}\sum_{-\infty<n<\infty}\oint z^{k-n-h}dz\phi_{n}\rangle\\
 & = & \langle\phi_{k+1}\sum_{-\infty<n<\infty}\frac{r^{k-n-h+1}\left(e^{2\pi i(k-n-h+1)}-1\right)}{i(k-n-h+1)}\phi_{n}\rangle\\
 & = & \frac{i\left(e^{-2\pi hi}-1\right)}{hr^{h}}
\end{eqnarray*}
where $r$ is the radius of the circular loop around the origin. So
we have constructed an observable which vanishes for highest weight
CFT but does not vanish for non-highest weight CFT. 

Another way to distinguish them is to compute the correlation function
in $l,r$ basis
\begin{eqnarray*}
\langle\left(L_{0}O(l,r)\right)\left(L_{0}O(l,r)\right)\rangle_{\text{{principal-series}}} & = & \langle l,r|L_{0}^{\dagger}L_{0}|l.r\rangle=l^{2}\\
\langle\left(L_{0}O(l,r)\right)\left(L_{0}O(l,r)\right)\rangle_{\text{{highest-weight}}} & = & \langle l-h,r-h|L_{0}^{\dagger}L_{0}|l-h.r-h\rangle=(l-h)^{2}\,.
\end{eqnarray*}
For principal series, we get integer squared and is independent of
the weight. Whereas for highest weight, it is non-integer and depends
on the weight of the operator. 

\section{OPE of principal series operators \label{sec:OPE-of-principal}}

In this section we derive operator product expansion (OPE) for the
principal series. First we will review the calculation for highest
weight CFT from \citep{Polyakov:1984yq}. Then we will extend the
derivation for principal series with suitable modification.

\subsection{Highest weight OPE}

Start with an ansatz
\begin{eqnarray}
O_{1}(z)O_{2}(0) & = & \sum_{k\geq0}\beta_{k}z^{\Delta_{3}-\Delta_{1}-\Delta_{2}+k}\left(\frac{\partial}{\partial\zeta}\right)^{k}O_{3}(\zeta)|_{\zeta\rightarrow0}\,.\label{eq:OPE ansatz}
\end{eqnarray}
Then using symmetry we can determine the coefficients. Commute left
side with $L_{1}$. Using the relation
\begin{eqnarray*}
[L_{1},O_{\Delta}(z)] & = & \left[z^{2}\frac{\partial}{\partial z}+2\Delta z\right]O_{\Delta}(z)
\end{eqnarray*}
we get
\begin{eqnarray*}
[L_{1},O_{1}(z)O_{2}(0)] & = & \left[z^{2}\frac{\partial}{\partial z}+2\Delta_{1}z\right]O_{1}(z)O_{2}(0)\,.
\end{eqnarray*}
Substituting the ansatz from equation \eqref{eq:OPE ansatz} in the
right side we get
\begin{eqnarray}
[L_{1},O_{1}(z)O_{2}(0)] & = & \sum_{k\geq0}\beta_{k}z^{\Delta_{3}-\Delta_{1}-\Delta_{2}+k+1}\left(\Delta_{3}+\Delta_{1}-\Delta_{2}+k\right)\left(\frac{\partial}{\partial\zeta}\right)^{k}O(\zeta)|_{\zeta\rightarrow0}\,.\label{eq:OPE lhs commute}
\end{eqnarray}
Now commuting $L_{1}$ with the right side of equation \eqref{eq:OPE ansatz}
we get
\begin{eqnarray}
\sum_{k\geq0}\beta_{k}z^{\Delta_{3}-\Delta_{1}-\Delta_{2}+k}\left(\frac{\partial}{\partial\zeta}\right)^{k}\left[L_{1},O(\zeta)\right]|_{\zeta\rightarrow0} & = & \sum_{k\geq0}\beta_{k}z^{\Delta_{3}-\Delta_{1}-\Delta_{2}+k}\left(\frac{\partial}{\partial\zeta}\right)^{k}\left(\zeta^{2}\frac{\partial}{\partial\zeta}+2\zeta\Delta_{3}\right)O(\zeta)|_{\zeta\rightarrow0}\,.\label{eq:OPE rhs commute}
\end{eqnarray}
We can now match the coefficient of power series of equations \eqref{eq:OPE lhs commute}
and \eqref{eq:OPE rhs commute}. Let us set $\Delta_{1}=\Delta_{2}$
for simplicity. As an example, let us match the coefficient of $z^{\Delta_{3}-\Delta_{1}-\Delta_{2}+1}$
\begin{eqnarray}
\beta_{0}\frac{\Gamma\left(\Delta_{3}+\Delta_{1}-\Delta_{2}+1\right)}{\Gamma\left(\Delta_{3}+\Delta_{1}-\Delta_{2}\right)}O(\zeta)|_{\zeta\rightarrow0} & = & \beta_{1}\left(\frac{\partial}{\partial\zeta}\right)\left(\zeta^{2}\frac{\partial}{\partial\zeta}+2\zeta\Delta_{3}\right)O(\zeta)|_{\zeta\rightarrow0}\nonumber \\
\beta_{0}\Delta_{3}O(\zeta)|_{\zeta\rightarrow0} & = & \beta_{1}\left(2\Delta_{3}+2\zeta\frac{\partial}{\partial\zeta}+\zeta^{2}\left(\frac{\partial}{\partial\zeta}\right)^{2}\right)O(\zeta)|_{\zeta\rightarrow0}\,.\label{eq:beta1}
\end{eqnarray}
For highest weight $O(\zeta)|_{\zeta\rightarrow0}$ is finite and
$\zeta\frac{\partial}{\partial\zeta}O(\zeta)|_{\zeta\rightarrow0}=0$.
Thus we get
\begin{eqnarray*}
\beta_{1} & = & \frac{\beta_{0}}{2}\,.
\end{eqnarray*}
Similarly matching all the terms, we get 
\begin{eqnarray}
O_{1}(z)O_{2}(0) & = & \beta_{123}\sum z^{\Delta_{3}-2\Delta_{1}}\ _{1}F_{1}\left(\Delta_{3},2\Delta_{3},z\frac{\partial}{\partial\zeta}\right)O_{3}(\zeta)|_{\zeta\rightarrow0}\,.\label{eq:OPE ansatz-1}
\end{eqnarray}
The above equality can also be derived, starting from the 3 point
function
\begin{eqnarray}
\langle O(z_{1})O(z_{2}\rightarrow0)O(z_{3})\rangle & =\frac{\beta_{123}}{z_{1}^{h}\left(z_{3}-z_{1}\right)^{h}z_{3}^{h}}= & \beta_{123}z_{1}^{-h}\left(1+h\frac{z_{1}}{z_{3}}+...\right)\frac{1}{z_{3}^{2h}}\,.\label{eq:3pt function h.w}
\end{eqnarray}

\subsection{Principal series}

For the principal series the OPE will take the form 
\begin{eqnarray}
O_{1}(z)O_{2}(0) & =\sum_{k>0}\beta_{-k}z^{\Delta_{3}-\Delta_{1}-\Delta_{2}-k}\left(L_{1}\right)^{k}O_{3}(\zeta)|_{\zeta\rightarrow0}+ & \sum_{k\geq0}\beta_{k}z^{\Delta_{3}-\Delta_{1}-\Delta_{2}+k}\left(L_{-1}\right)^{k}O_{3}(\zeta)|_{\zeta\rightarrow0}\,.\label{eq:OPE ansatz ps}
\end{eqnarray}
Here we have also added terms with $L_{1}O$ because for principal
series $L_{1}O\neq0$ in general. Again we commute with $L_{1}$ to
determine $\beta_{k}$. Commuting the left side of equation \eqref{eq:OPE ansatz ps}
we get
\begin{eqnarray*}
[L_{1},O_{1}(z)O_{2}(0)] & = & \sum_{k>0}\beta_{k}z^{\Delta_{3}-\Delta_{1}-\Delta_{2}+k}\left(z\left(\Delta_{3}+\Delta_{1}-\Delta_{2}+k\right)\left(L_{-1}\right)^{k}+\left(\zeta^{2}\frac{\partial}{\partial\zeta}+2\Delta_{2}\zeta\right)\left(L_{-1}\right)^{k}\right)O(\zeta)\,.
\end{eqnarray*}
Commuting the right side of equation \eqref{eq:OPE ansatz ps} we
get 
\begin{eqnarray*}
[L_{1},O_{1}(z)O_{2}(0)] & = & \sum_{k>0}\beta_{k}z^{\Delta_{3}-\Delta_{1}-\Delta_{2}+k}\left[L_{1},\left(L_{-1}\right)^{k}O(\zeta)\right]\,.
\end{eqnarray*}
Equating the above two equations gives
\begin{eqnarray}
 &  & \beta_{k+1}\left(L_{-1}^{k+1}\left(\zeta^{2}\frac{\partial}{\partial\zeta}+2\Delta_{3}\zeta\right)-\left(\zeta^{2}\frac{\partial}{\partial\zeta}+2\Delta_{2}\zeta\right)\left(L_{-1}\right)^{k+1}\right)O(\zeta)\nonumber \\
 & = & \beta_{k}\left(\Delta_{3}+\Delta_{1}-\Delta_{2}+k\right)\left(L_{-1}\right)^{k}O(\zeta)\,.\label{eq:betak+}
\end{eqnarray}

Similarly, to determine $\beta_{-k}$ we can commute both sides with
$L_{-1}$ 
\begin{eqnarray*}
[L_{-1},O_{1}(z)O_{2}(0)] & = & \sum_{k>0}\beta_{-k}z^{\Delta_{3}-\Delta_{1}-\Delta_{2}-k-1}\left(\left(\Delta_{3}-\Delta_{1}-\Delta_{2}-k\right)\left(L_{1}\right)^{k}+z\left(L_{1}\right)^{k}L_{-1}\right)O(\zeta)\,.
\end{eqnarray*}
Commuting the right side of equation \eqref{eq:OPE ansatz ps} we
get 
\begin{eqnarray*}
[L_{-1},O_{1}(z)O_{2}(0)] & = & \sum_{k<0}\beta_{-k}z^{\Delta_{3}-\Delta_{1}-\Delta_{2}-k}\left[L_{-1},\left(L_{1}\right)^{k}O(\zeta)\right]\,.
\end{eqnarray*}
Equating the two sides we get
\begin{eqnarray}
\beta_{-k-1}\left[L_{-1},L_{1}^{k+1}\right]O(\zeta)|_{\zeta\rightarrow0} & = & \beta_{-k}\left(\Delta_{3}-\Delta_{1}-\Delta_{2}-k\right)\left(L_{1}\right)^{k}O(\zeta)\,.\label{eq:betak-}
\end{eqnarray}
Simplifying equation \eqref{eq:betak+} gives
\begin{eqnarray}
 &  & \beta_{k+1}\left(2\left(k+1+\Delta_{3}-\Delta_{2}\right)\zeta\left(\frac{\partial}{\partial\zeta}\right)^{k+1}+\left(k+1\right)\left(k+2\Delta_{3}\right)\left(\frac{\partial}{\partial\zeta}\right)^{k}\right)O(\zeta)\nonumber \\
 & = & \beta_{k}\left(\Delta_{3}+\Delta_{1}-\Delta_{2}+k\right)\left(\frac{\partial}{\partial\zeta}\right)^{k}O(\zeta)\,.\label{eq:beta+recursion}
\end{eqnarray}

An important thing to note is that recursion relations explicitly
depend on $O(\zeta)$. Now we substitute the expansion 
\begin{eqnarray*}
O_{3}(\zeta) & = & \sum_{j}\frac{O_{3j}}{\zeta^{h+j}}
\end{eqnarray*}
and compare the coefficient of same power of $\zeta$, we get
\begin{eqnarray*}
\beta_{k+1} & = & \beta_{k}\frac{\left(\Delta_{3}+\Delta_{1}-\Delta_{2}+k\right)}{\left(2\left(k+1+\Delta_{3}-\Delta_{2}\right)(-h-j-k)+\left(k+1\right)\left(k+2\Delta_{3}\right)\right)}\,.
\end{eqnarray*}
We find that $\beta_{k}$ depends on $j$. This suggests that we must
start with an OPE of the form 
\begin{eqnarray}
O_{1}(z)O_{2}(\zeta) & = & \sum_{j}\left(\sum_{k>0}\beta_{-k,j}z^{\Delta_{3}-\Delta_{1}-\Delta_{2}-k}\left(L_{1}\right)^{k}\frac{O_{3j}}{\zeta^{j}}+\sum_{k>0}\beta_{k,j}z^{\Delta_{3}-\Delta_{1}-\Delta_{2}+k}\left(L_{-1}\right)^{k}\frac{O_{3j}}{\zeta^{j}}\right)\,.\label{eq:ope-mode}
\end{eqnarray}
Then going through the above derivation we get
\begin{eqnarray*}
\beta_{k+1,j} & = & -\beta_{k,j}\frac{\left(\Delta_{3}+\Delta_{1}-\Delta_{2}+k\right)}{\left(k-K_{+}\right)(k-K_{-})}\\
\beta_{-k-1,j} & = & \beta_{-k,j}\frac{\left(\Delta_{3}-\Delta_{1}-\Delta_{2}-k\right)}{(k+2\Delta_{3}+2j+2h)(k+1)}
\end{eqnarray*}
where 
\begin{eqnarray*}
K_{\pm} & = & \frac{1}{2}\left(1+2\left(\Delta_{2}-j-h\right)\pm\sqrt{\left(1+\Delta_{2}\right)^{2}+4\left(2\Delta_{3}-j-h\right)\left(1-j-h\right)}\right)\,.
\end{eqnarray*}
Calculations are shown in the appendix.

Then equation \eqref{eq:ope-mode} can be written in terms of hypergeometric
functions
\begin{eqnarray*}
O_{1}(z)O_{2}(\zeta) & = & \sum_{j}z^{\Delta_{3}-\Delta_{1}-\Delta_{2}}\beta_{0,j}\Bigg(\ _{1}F_{1}\left(\Delta_{1}+\Delta_{2}-\Delta_{3};2\Delta_{3}+2j+2h;-\frac{1}{z}\left(\zeta^{2}\frac{\partial}{\partial\zeta}+2\Delta_{3}\zeta\right)\right)\frac{O_{3j}}{\zeta^{h+j}}\\
 &  & +\ _{2}F_{2}\left(\Delta_{3}+\Delta_{1}-\Delta_{2},1;K_{+},K_{-};-z\frac{\partial}{\partial\zeta}\right)\frac{O_{3j}}{\zeta^{h+j}}\Bigg)\,.
\end{eqnarray*}
This is the conformal partial wave expansion for the principal series. 

Using equation \eqref{eq:z-l-k}, we can show that above equation
is equivalent to three point function
\begin{eqnarray*}
\langle O(z_{1})O(z_{2}\rightarrow0)O(z_{3})\rangle & = & \frac{\beta_{123}}{z_{1}^{h}\left(z_{3}-z_{1}\right)^{h}z_{3}^{h}}
\end{eqnarray*}
given $\beta_{0,j}=\beta_{123}$. 

The main conclusion is that there are infinitely many singular terms
coming from terms like $L_{1}^{k}O$ in the OPE. The OPE therefore
has an essential singularity, unlike any known conformal field theory
that may be viewed as arising from a renormalizable field theory.
This puts the set of interacting conformal field theories based on
representations containing the principal series well outside the class
of conventional quantum field theories. The OPE also depends on an
infinite number of parameters that are free at this level of analysis,
compared to the single parameter one normally encounters in CFT. If
these CFTs of relevance for de Sitter space exist, it seems they have
more in common with non-renormalizable theories, than with conventional
CFTs.

Finally we note all the conclusions of the present section carry over
to the complementary series, provided we take $h$ in the appropriate
range $1>h>1/2$.

\section{Conclusion}

de-Sitter holography implies that bulk and boundary states should
be in principal, complementary, discrete series and indecomposible
representations. Some of the details of these representations were
studied from the conformal field theory perspective. In particular,
we analyzed the implications of global conformal invariance for the
operator product expansion. Because the weights of the principal and
complementary series are unbounded, there end up being infinitely
many singular terms in the operator product expansion. Nevertheless,
this is compatible with the usual simplifications of the two and three-point
functions of quasi-primary operators. The essential singularity present
in these operator product expansions is not reproducible from conventional
quantum field theories.
\begin{acknowledgments}
This research was supported in part by DOE grant de-sc0010010.
\end{acknowledgments}

\section*{Appendix}

\subsection{Bulk isometries}

Bulk $SO(3,1)$ isometries can be expressed in terms of embedding
coordinates $X_{A}=(Y_{1},Y_{2},Z,T)$
\begin{eqnarray}
z_{AB} & = & i\left(X_{B}\partial_{A}-X_{A}\partial_{B}\right)\label{eq:bulkdiff}
\end{eqnarray}
where de Sitter spacetime is the hyperboloid
\[
R^{2}=Y_{1}^{2}+Y_{2}^{2}+Z^{2}-T^{2}\,.
\]
Poincare coordinates $(y_{1},y_{2},\eta)$ are given by 
\begin{eqnarray*}
T & = & \frac{R}{2}(\eta-\frac{1}{\eta})-\frac{1}{2R\eta}(y_{1}^{2}+y_{2}^{2})\\
Y_{1} & = & \frac{y_{1}}{\eta}\\
Y_{2} & = & \frac{y_{2}}{\eta}\\
Z & = & \frac{R}{2}(\eta+\frac{1}{\eta})-\frac{1}{2R\eta}(y_{1}^{2}+y_{2}^{2})\,.
\end{eqnarray*}
With inverse relations
\begin{eqnarray*}
\eta & = & \frac{\sqrt{Y_{1}^{2}+Y_{2}^{2}+Z^{2}-T^{2}}}{Z-T}\\
y_{1} & = & \frac{Y_{1}}{Z-T}\\
y_{2} & = & \frac{Y_{2}}{Z-T}\,.
\end{eqnarray*}

Equation \eqref{eq:bulkdiff} in $R,\eta,y_{1},y_{2}$ coordinates
becomes
\begin{eqnarray*}
J_{3}\equiv J_{Y_{1}Y_{2}} & = & i\left(y_{2}\partial_{y_{1}}-y_{1}\partial_{y_{2}}\right)\\
J_{2}\equiv J_{ZY_{1}} & = & -i\left(\frac{1+y_{1}^{2}-y_{2}^{2}+\eta^{2}}{2}\partial_{y_{1}}+y_{1}y_{2}\partial_{y_{2}}+y_{1}\eta\partial_{\eta}\right)\\
-J_{1}\equiv J_{ZY_{2}} & = & -i\left(\frac{1-y_{1}^{2}+y_{2}^{2}+\eta^{2}}{2}\partial_{y_{2}}+y_{1}y_{2}\partial_{y_{1}}+y_{2}\eta\partial_{\eta}\right)\\
K_{1}\equiv K_{Y_{1}T} & = & -i\left(\frac{-1+y_{1}^{2}-y_{2}^{2}+\eta^{2}}{2}\partial_{y_{1}}+y_{1}y_{2}\partial_{y_{2}}+y_{1}\eta\partial_{\eta}\right)\\
K_{2}\equiv K_{Y_{2}T} & = & -i\left(\frac{-1-y_{1}^{2}+y_{2}^{2}+\eta^{2}}{2}\partial_{y_{2}}+y_{1}y_{2}\partial_{y_{1}}+y_{2}\eta\partial_{\eta}\right)\\
K_{3}\equiv K_{ZT} & = & -i\left(y_{1}\partial_{y_{1}}+y_{2}\partial_{y_{2}}+\eta\partial_{\eta}\right)\,.
\end{eqnarray*}
We can go to the complex coordinate $z=y_{1}+iy_{2}$ and define
\begin{eqnarray*}
J_{L}=z\partial_{z}+\frac{\eta}{2}\partial_{\eta}\ \ \ ,\ \ \  & K_{+L}=i\left(z^{2}\partial_{z}+\eta^{2}\partial_{\bar{z}}+z\eta\partial_{\eta}\right)\ \ \ ,\ \ \  & K_{-L}=-i\partial_{z}\\
J_{R}=-\bar{z}\partial_{\bar{z}}-\frac{\eta}{2}\partial_{\eta}\ \ \ ,\ \ \  & K_{-R}=-i\left(\bar{z}^{2}\partial_{\bar{z}}+\eta^{2}\partial_{z}+\bar{z}\eta\partial_{\eta}\right)\ \ \ ,\ \ \  & K_{+R}=i\partial_{\bar{z}}\,.
\end{eqnarray*}
We see that they take very simple form in Poincare coordinates compared
to spherical coordinates. 

\subsection{OPE calculation for principal series}

Calculation of $\beta_{k,j}$ is same as in equation \eqref{eq:beta+recursion}.
\begin{eqnarray*}
 &  & \beta_{k+1,j}\left(2\left(k+1+\Delta_{3}-\Delta_{2}\right)\zeta\left(\frac{\partial}{\partial\zeta}\right)^{k+1}+\left(k+1\right)\left(k+2\Delta_{3}\right)\left(\frac{\partial}{\partial\zeta}\right)^{k}\right)\zeta^{-h-j}\\
 & = & \beta_{k,j}\left(\Delta_{3}+\Delta_{1}-\Delta_{2}+k\right)\left(\frac{\partial}{\partial\zeta}\right)^{k}\zeta^{-h-j}\\
 &  & \beta_{k+1,j}\left(2\left(k+1+\Delta_{3}-\Delta_{2}\right)(-h-j)...(-h-j-k)+\left(k+1\right)\left(k+2\Delta_{3}\right)(-h-j)...(-h-j-k+1)\right)\zeta^{-h-j-k}\\
 & = & \beta_{k,j}\left(\Delta_{3}+\Delta_{1}-\Delta_{2}+k\right)(-h-j)...(-h-j-k+1)\zeta^{-h-j-k}\\
 &  & \beta_{k+1,j}\left(2\left(k+1+\Delta_{3}-\Delta_{2}\right)(-j-h-k)+\left(k+1\right)\left(k+2\Delta_{3}\right)\right)\zeta^{-h-j-k}\\
 & = & \beta_{k,j}\left(\Delta_{3}+\Delta_{1}-\Delta_{2}+k\right)\zeta^{-h-j-k}\\
 &  & \beta_{k+1,j}\\
 & = & -\beta_{k,j}\frac{\left(\Delta_{3}+\Delta_{1}-\Delta_{2}+k\right)}{\left(k-K_{+}\right)(k-K_{-})}
\end{eqnarray*}

where
\begin{eqnarray*}
K_{\pm} & = & \frac{1}{2}\left(1+2\left(\Delta_{2}-j-h\right)\pm\sqrt{\left(1+\Delta_{2}\right)^{2}+4\left(2\Delta_{3}-j-h\right)\left(1-j-h\right)}\right)\,.
\end{eqnarray*}
Calculation of $\beta_{-k,j}$ is 
\begin{eqnarray*}
 &  & \beta_{-k-1,j}\left[L_{-1},L_{1}^{k+1}\right]\zeta^{-h-j}\\
 & = & \beta_{-k,j}\left(\Delta_{3}-\Delta_{1}-\Delta_{2}-k\right)\left(L_{1}\right)^{k}\zeta^{-h-j}\\
 &  & \beta_{-k-1,j}\left(\left(2\Delta_{3}+k-h-j-1\right)...\left(2\Delta_{3}-h-j-1\right)\left(-h-j\right)-\left(2\Delta_{3}+k-h-j\right)...\left(2\Delta_{3}-h-j\right)\left(k-h-j+1\right)\right)\zeta^{-h-j+k}\\
 & = & \beta_{-k,j}\left(\Delta_{3}-\Delta_{1}-\Delta_{2}-k\right)\left(2\Delta_{3}+k-1-h-j\right)...\left(2\Delta_{3}-h-j\right)\zeta^{-h-j+k}\\
 &  & \beta_{-k-1,j}\left(\left(2\Delta_{3}-h-j-1\right)\left(-h-j\right)-\left(2\Delta_{3}+k-h-j\right)\left(k-h-j+1\right)\right)\zeta^{-h-j+k}\\
 & = & \beta_{-k,j}\left(\Delta_{3}-\Delta_{1}-\Delta_{2}-k\right)\zeta^{-h-j+k}\\
 &  & \beta_{-k-1,j}\\
 & = & \beta_{-k,j}\frac{\left(\Delta_{3}-\Delta_{1}-\Delta_{2}-k\right)}{(k+2\Delta_{3}+2j+2h)(k+1)}\,.
\end{eqnarray*}

\bibliographystyle{utphys}
\bibliography{ds_ope}

\end{document}